\documentclass[pra,twocolumn, aps, preprintnumbers,superscriptaddress]{revtex4-2}
\usepackage[colorlinks=true,citecolor=blue,urlcolor=black]{hyperref}
\usepackage{verbatim}
\usepackage{amsmath}
\usepackage{latexsym}
\usepackage{revsymb}
\usepackage{yfonts}
\usepackage{ifthen}
\usepackage{url}
\usepackage{natbib}
\usepackage{amsfonts}
\usepackage{amsmath}
\usepackage{amssymb}
\usepackage{amsthm}
\usepackage{graphicx}
\usepackage{bm}
\usepackage{bbm}
\usepackage{epsfig,color,amssymb}
\usepackage{subfigure}
\usepackage{amsfonts}
\usepackage{amscd}
\usepackage{amsmath}
\usepackage{multirow}
\usepackage{chemarrow}
\usepackage{dcolumn}
\usepackage{bm}
\usepackage{graphicx}
\usepackage{enumerate}
\usepackage{epsfig}
\usepackage{subfigure}
\usepackage{xcolor}
\usepackage{multirow}
\usepackage{ulem}
\usepackage{braket}
\usepackage{comment}
\usepackage{enumitem}
\usepackage{amsthm}
\renewcommand{\emph}[1]{\textit{#1}}
\def\fid#1#2{\left\langle{#1}|{#2}\right\rangle}

\begin{document}
	
\title{Experimental quantum e-commerce}

\author{Xiao-Yu Cao}\thanks{These authors contributed equally.}
\author{Bing-Hong Li}\thanks{These authors contributed equally.}
\affiliation{National Laboratory of Solid State Microstructures and School of Physics, Collaborative Innovation Center of Advanced Microstructures, Nanjing University, Nanjing 210093, China}
\affiliation{Department of Physics and
Beijing Key Laboratory of Opto-electronic Functional Materials and Micro-nano Devices,
Key Laboratory of Quantum State Construction and Manipulation (Ministry of Education),
Renmin University of China, Beijing 100872, China}

  \author{Yang Wang}
        \affiliation{National Laboratory of Solid State Microstructures and School of Physics, Collaborative Innovation Center of Advanced Microstructures, Nanjing University, Nanjing 210093, China}
        \affiliation{Henan Key Laboratory of Quantum Information and Cryptography, SSF IEU, Zhengzhou 450001, China}
\author{Yao Fu}
\affiliation{Beijing National Laboratory for Condensed Matter Physics and Institute of Physics, Chinese Academy of Sciences, Beijing 100190, China}
\author{Hua-Lei Yin}\email{hlyin@ruc.edu.cn}
\affiliation{Department of Physics and
Beijing Key Laboratory of Opto-electronic Functional Materials and Micro-nano Devices,
Key Laboratory of Quantum State Construction and Manipulation (Ministry of Education),
Renmin University of China, Beijing 100872, China}
\affiliation{National Laboratory of Solid State Microstructures and School of Physics, Collaborative Innovation Center of Advanced Microstructures, Nanjing University, Nanjing 210093, China}
\author{Zeng-Bing Chen}\email{zbchen@nju.edu.cn}
\affiliation{National Laboratory of Solid State Microstructures and School of Physics, Collaborative Innovation Center of Advanced Microstructures, Nanjing University, Nanjing 210093, China}
\affiliation{MatricTime Digital Technology Co. Ltd., Nanjing 211899, China}
	
	
	\begin{abstract}
E-commerce, a type of trading that occurs at a high frequency on the Internet, requires guaranteeing the integrity, authentication and non-repudiation of messages through long distance. As current e-commerce schemes are vulnerable to computational attacks, quantum cryptography, ensuring information-theoretic security against adversary's repudiation and forgery, provides a solution to this problem. However, quantum solutions generally have much lower performance compared to classical ones. Besides, when considering imperfect devices, the performance of quantum schemes exhibits a notable decline.
Here, we demonstrate the whole e-commerce process of involving the signing of a contract and payment among three parties by proposing a quantum e-commerce scheme, which shows resistance of attacks from imperfect devices. Results show that with a maximum attenuation of 25 dB among participants, our scheme can achieve a signature rate of 0.82 times per second for an agreement size of approximately 0.428 megabit. This proposed scheme presents a promising solution for providing information-theoretic security for e-commerce.
	\end{abstract}

	\maketitle
\section{introduction}
Developing algorithms and quantum attacks threaten the security of classic cryptography~\cite{shor1999polynomial,martin2012experimental,boudot2020comparing,gouzien2021factoring}. Since the security of current
cryptographic schemes tend to rely on computationally hard mathematical problems~\cite{rivest1978method,demillo1978foundations,elgamal1985public,silverman1992rational}, information-theoretic security against unlimited computational power, has been a hot topic.
Quantum mechanic laws is one of the approaches~\cite{bennett2014quantum,ekert1991quantum,Yin:2017:Experimental,schiansky2023demonstration,Zhou2023exp,yinnwac228}. Quantum key distribution (QKD), which is the most mature application field of quantum technology, offers two remote users unconditionally secure keys~\cite{bennett2014quantum,ekert1991quantum}. Combined with one-time pad, QKD successfully guarantees the confidentiality of messages. 

Secure identification is another important application domain in the realm of the quantum internet~\cite{wehner2018quantum}, which can help guarantee the security during financial transactions.
E-commerce, as an indispensable part of daily life, requires identification of the parties and non-repudiation of the contract. A commitment among different participants is required to guarantee the validity of a transaction. The security of classical e-commerce schemes tends to be based on public-key cryptography algorithms~\cite{rivest1978method,demillo1978foundations,elgamal1985public,silverman1992rational}, which is only secure assuming the limited computation power, and there lacks an effective solution to defend against external attacks. Additionally, the presence of dishonest participants may render the contract invalid.
Cryptography contains four main information security objectives, confidentiality, integrity, authenticity, and non-repudiation~\cite{menezes2018handbook}. The integrity, authenticity, and non-repudiation of messages need to be assured in an e-commerce scheme. The integration of QKD with one-time pad, only promising confidentiality for messages, fails to accomplish this task. Quantum digital signatures (QDS) can provide information-theoretic security for the last three primitives, and thus is suitable for e-commerce scenarios.

QDS was proposed first at 2001~\cite{gottesman2001quantum}. The original version has many impractical experimental requirements, which made its implementation impossible with available technology. With the development in the next decade, the requirements of swap test and quantum memory are removed~\cite{clarke2012experimental,dunjko2014quantum,PhysRevLett.113.040502}. Nevertheless, their security analyses were based on secure quantum channels. In 2016, two schemes were proposed independently to solve this problem~\cite{yin2016practical,amiri2016secure}. Triggered by the two protocols and developments in QKD~\cite{lo2012measurement,yin2016MDI404,lucamarini2018overcoming,PhysRevLett.123.100506,cz-19,WangYin-20,zhou_22,xie2022breaking,zeng2022mode,Xie2023Scalable,Xie2023Advantages}, many achievements have been made theoretically~\cite{puthoor2016measurement,PhysRevA.94.042314,Thornton:2019:CV,lu2021efficient,zhang2021twin,Weng:21,qin2022quantum} and experimentally~\cite{Collins:16, Yin:2017:Exp,roberts2017experimental, PhysRevApplied.10.034033,An:19,Ding:20, Richter2021Agile, pelet2022unconditionally}.

Previous QDS schemes are inefficient when it comes to multi-bit cases and their performances are far from classical solutions. Recently, a new scheme, based on secret sharing, one-time pad, and one-time universal hashing (OTUH), has been proposed~\cite{yinnwac228}. This scheme can sign an arbitrarily long document with a relatively short signature, whose performance outperforms all previous protocols. Furthermore, a variant of this scheme reduces the requirements on keys~\cite{li2023one}, in which the privacy amplification steps are removed. Besides, security proof of previous schemes tends to be based on assumptions on the ideal devices and there has been a lack of QDS schemes with the ability to solve the loopholes from imperfect devices.
To enhance the roubstness to imperfect devices, our key generation process (KGP) draws on the development of QKD~\cite{lo2012measurement,Braunstein:2012:Side-Channel-Free, PhysRevA.90.052314,PhysRevA.92.032305,Tang2016Exp, yoshino2018quantum, pereira2019quantum,PhysRevApplied.15.034072,PhysRevApplied.12.054034,Xu2020Secureqkd,Pereiraeaaz4487,zhang2022experimentals,GU2022,fan2022robust,fan2021measurement}. 
Four-phase measurement-device-independent (MDI) QKD~\cite{GU2022} is secure against possible source flaws and outperforms other protocols in key rate and KGP of our scheme based on this protocol retains its advantage.

Here, we present a quantum solution for e-commerce scenarios by proposing an efficient quantum e-commerce scheme based on the ideas mentioned above, which offers security advantages over classical schemes.
Motivated by OTUH scheme~\cite{yinnwac228, li2023one} and four-phase MDI-QKD~\cite{GU2022}, our scheme is able to sign multi-bit documents with high efficiency while mitigating the impact of imperfect devices, and thus improves the overall security and practicality of the scheme. 
The signature rate is only limited by the minimum key rate of KGP between different participants and merchant.
Our experimental implementation on a multi-user quantum network successfully signs a 0.428-megabit (Mb) agreement 0.82 times per second with a maximum attenuation of 25 dB between a participant and merchant.
We also characterize the imperfections of the sources experimentally. Our work contributes to the further development of e-commerce in the quantum era by providing a practical and efficient solution with enhanced security.

\begin{figure*}
	\centering
	\includegraphics[width=\textwidth]{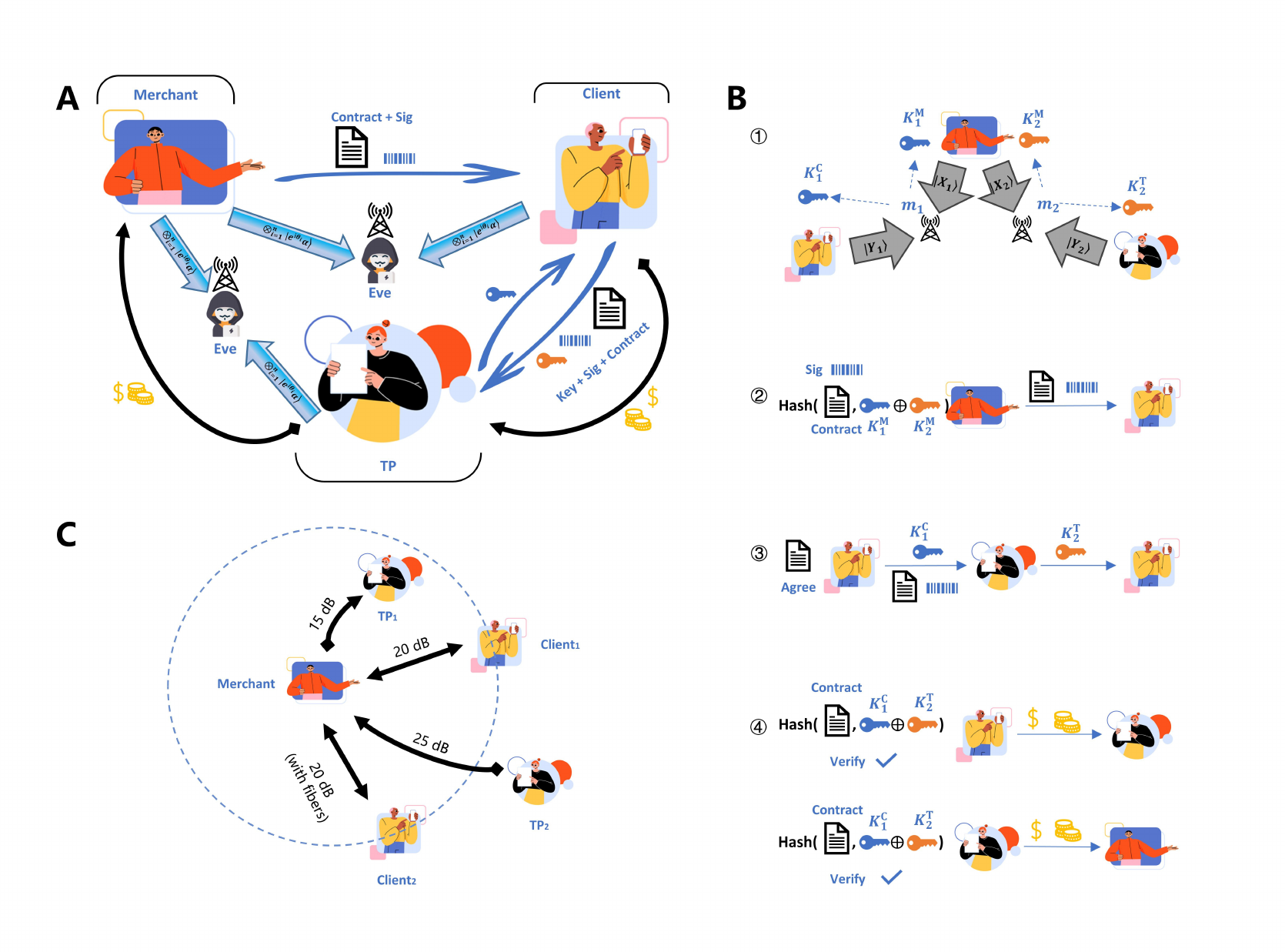}\\
	\caption{\textbf{Illustrations depicting the process and network of quantum e-commerce. (A)} Illustration of the process of quantum e-commerce. We consider the three-party scenario where Client buys a product from Merchant. TP is introduced as an arbiter to prevent either Merchant or Client from cheating. Merchant shares two sequences of coherent quantum states with Clinet and TP, respectively. Merchant then generates the contract with all information of the e-commerce, and obtains a signature through a hash function and keys distilled by his sequences. Thereafter, Merchant sends the contract and signature to Client. Client, if agreeing with the contract, will send the contract, signature and keys distilled by his consequence to TP. TP will then send keys distilled by his own sequence back to Client. Both Client and TP independently verify the signature through their own and received keys by hash functions. Client will pay the money to TP if he verifies the signature. TP will transfer the money to Merchant if he also passes the signature. \textcolor{black}{\textbf{(B)} A flow chart of the protocol. Details of the procedure are explained in protocol description step by step.} \textbf{(C)} A diagram of users in quantum networks.
    }\label{f1}
\end{figure*}

\section{Protocol description}\label{sec2}
In a classical e-commerce scenario, a third party (TP) is always required and assumed as trusted. Thus, only those with high authority can be a TP, leading to a requirement of centralized systems. Here, we propose a three-party quantum e-commerce protocol with no assumptions on TP.
\textcolor{black}{The only requirement is that  the majority of the three parties must be honest, which is a basic requirement of all three-party protocols.}
In the protocol we describe a scenario where Client buys a product from Merchant through a network, and TP is an arbiter to help finish the whole process successfully. 
In the network with numerous nodes, TP can be an arbitrary party since there are no assumptions on it. A decentralized system can be realized based on our proposed three-party protocol. \textcolor{black}{A schematic of the quantum e-commerce protocol is shown in Fig.~\ref{f1} (A). The step-by-step procedure is comprehensively elucidated below and shown in Fig.~\ref{f1} (B).}  

(i) Distribution: \textcolor{black}{Four-phase MDI-KGP is utilized here to generate raw keys and the form of quantum states is $\otimes_{i=1}^n\ket{e^{{\rm i}\theta_i}\alpha}$, where $\theta_i$ is the modulated phase and $|\alpha|^2$ is the intensity of pulses. Details about KGP can be seen in Materials and Methods.} Merchant prepares two sequences of coherent states $\ket{\bf{X_1}}$ and $\ket{\bf{X_2}}$, and keeps the phase of every state.
Client also prepares a sequence $\ket{\bf{Y_1}}$ and the TP prepares $\ket{\bf{Y_2}}$. Through a Merchant-Client quantum channel, Merchant sends $\ket{\bf{X_1}}$ and Client sends $\ket{\bf{Y_1}}$ to an untrusted intermediate Eve who performs interference measurements on the received pulses with a 50: 50 beam splitter (BS) and two single-photon detectors, and announces the detection results $\bf{m_1}$. 
Likewise,  $\ket{\bf{X_2}}$ and $\ket{\bf{Y_2}}$ are sent to an untrusted intermediate through the Merchant-TP channel, and the detection results $\bf{m_2}$ are announced.


(ii) Signature: To sign a contract, Merchant distills a $3n$-bit key $\bf{k_1^{\rm{M}}}$ from $\ket{\bf{X_1}}$ and $\bf{m_1}$. He will select states in $\ket{\bf{X_1}}$ that successfully intervene with that in $\ket{\bf{Y_1}}$ according to the detection results $m_1$. He also communicates with Client through an authenticated channel to sift the keys, and finally chooses a $3n$-bit substring to form $\bf{k_1^{\rm{M}}}$.
Likewise, Merchant distills a $3n$-bit key $\bf{k_2^{\rm{M}}}$ from $\ket{\bf{X_2}}$ and $\bf{m_2}$.

The keys $\bf{k_1^{\rm{M}}} \stackrel{m_{1}}{\longleftarrow} \ket{\bf{X_1}}$ and $\bf{k_2^{\rm{M}}} \stackrel{\bf{m_{2}}}{\longleftarrow} \ket{\bf{X_2}}$ are used to generate the signature of the contract. 
The signature is generated through a universal hash function $\bf{Sig}$$=Hash(\bf{C},\bf{k_1^{\rm{M}}}\oplus \bf{k_2^{\rm{M}}})$, with a length of $n$-bit, where $\bf{C}$ is the contract containing all details including timestamp, the identity of Merchant and Client. The function $Hash$ is composed of a linear feedback shift register (LFSR) Toeplitz functions.  Details are shown in Materials and Methods.

(iii) Transference: Merchant sends the contract and signature $\{\bf{C},\bf{Sig}\}$ to Client. If Client agrees with the contract, he distills a key $\bf{k_1^{\rm{C}}} \stackrel{\bf{m_{1}}}{\longleftarrow} \ket{\bf{Y_1}}$, following the rules same as that of Merchant. He then send $\{\bf{C},\bf{Sig},\bf{k_1^{\rm{C}}}\}$ to TP.

TP also obtains $\bf{k_2^{\rm{T}}}\stackrel{m_{2}}{\longleftarrow} \ket{\bf{Y_2}}$. He will send his key $\bf{k_2^{\rm{T}}}$ to Client after he receives $\{\bf{C},\bf{Sig},\bf{k_1^{\rm{C}}}\}$.

(iv) Verification and payment: Both Client and TP independently verify the signature by calculating $Hash(\bf{C},\bf{k_1^{\rm{C}}}\oplus \bf{k_2^{\rm{T}}})$ and comparing the result with $\bf{Sig}$.
If the result is identical to $\bf{Sig}$, the signature is successfully passed.
Client will pay the money to TP if he verifies the signature.
TP, after receiving the payment, will transfer the money to Merchant if he passes the signature. Otherwise, he will return the money to Client and announce that the contract is aborted.

In the distribution step, the participants essentially share correlated and secret quantum states. The secrecy of the interpreted keys, together with one-time hashing, protects the security of the program against Client's tampering attacks. \textcolor{black}{Note that the distribution step is different from QKD. Privacy amplification is removed because secrecy leakage of keys can be tolerated in signature tasks through one-time universal hashing~\cite{li2023one}.}
The transference step guarantees that Client and TP obtain the same final keys. Merchant's repudiation attacks are prevented by TP because TP will always make the correct judgment if Client is honest.

In a traditional digital payment scheme, the third party is assumed trusted and corresponding to a central authority. \textcolor{black}{Previous quantum-digital authentication/payment schemes follow this centralized structure where the security is guaranteed by the trusted central authority~\cite{schiansky2023demonstration}.} In the proposed scheme, the assumption on the third party is removed and the security is guaranteed by the majority of honest parties. Under this structure, the status of the three parties is equal and thus the third party can be decentralized in real implementations. Compared with classical solutions, the proposed quantum e-commerce requires additional quantum channels of Merchant-Client and Merchant-TP, while providing information-theoretic security and decentralized character, thus mitigating the burden of the authority and risk of insider attacks. The proposed scheme has great potential in a future block chain payment system.


\section{Security analysis} 
The proposed scheme is a three-party protocol without strong assumptions on TP. In other words, the three parties have equal status and the final decision is made by voting principle if disagreement happens. It must be assumed that at most one party can be malicious. Otherwise, malicious parties can cooperate to finish the attack by controlling the voting result. 
In security analysis, we consider four cases: honest abort, Merchant’s repudiation attack, Client's forgery attack, and TP's forgery attack.

\bigskip\noindent\textbf{\it{Robustness.}}
If Merchant and Client (or TP) share different key bits after the distribution stage, the protocol will be aborted even if the users are all honest. That is, an honest run abortion occurs. In the protocol, Merchant and Client (TP) perform error correction on their final keys, with a failure probability of no more than $\epsilon_{\rm{EC}}$.
The correctness of classical information transference is protected by classical information technology such as authenticated channels, whose failure probability is set as no more than $\epsilon'$.
The probability that Merchant and Client share different final key is no more than $\epsilon_{\rm{EC}}+\epsilon'$, and the same for Merchant and TP. 
Thus the robustness bound is $\epsilon_{\rm{rob}}=2\epsilon_{\rm{EC}}+2\epsilon'$.
Since $\epsilon'$ is a parameter of classical communication, we assume it as $\epsilon'=10^{-10}$ in the simulation.

\bigskip\noindent\textbf{\it{Repudiation.}}
In a repudiation attack, Merchant attempts to let Client accept the contract while TP rejects it so that he can successfully deny the contract. 
For Merchant’s repudiation attacks, Client and TP are both honest and symmetric and thus hold the same new key strings. They will make the same decision for the same contract and signature.
Repudiation attacks succeed only when errors occur in one of the transference steps. The repudiation bound is $\epsilon_{\rm{rep}}=2\epsilon'$.

\bigskip\noindent\textbf{\it{Forgery.}}
In Client's forgery attack, Client will tamper the contract and attempts to let TP accept the tampered contract forwarded to him. According to our protocol, TP accepts the contract if and only if TP obtains the same result as $\bf{Sig}$ through one-time hash functions. Actually, this is the same as an authentication scenario where Client is the attacker attempting to forge the information sent from Merchant to TP.
TP also has the motivation to perform forgery attack. He may attempt to intercept the contract $\bf{C}$ and  tamper the information of price in it. If Client agrees with the tampered (higher) price, TP can earn the price difference secretly. This is the same as an authentication scenario where TP is the attacker attempting to forge the information sent from Merchant to Client. Thus, Client and TP's forgery attacks are equivalent, and we only analyze Client's in the following.

In the hash function $Hash(C,\bf{k_1^{\rm{M}}}\oplus \bf{k_2^{\rm{M}}})$, $\bf{k_1^{\rm{M}}}\oplus \bf{k_2^{\rm{M}}}$ is actually divided into three $n$-bit substrings. Here we rewrite it as $\bf{k_1^{\rm{M}}}\oplus \bf{k_2^{\rm{M}}}=(x_2,x_3,x_4)$ to keep consistent with that in Materials and Methods.
Define  $\mathcal{H}_n=H_{\rm{min}}(\mathbf{X}|\mathbf{B})_\rho$ as the min-entropy of $\mathbf{X}$ and $\mathbf{B}$, where $\mathbf{X}\in \{\bf{x_2},\bf{x_3},\bf{x_4}\}$ and $\mathbf{B}$ represents Client's guessing for $\mathbf{X}$. We can estimate $\mathcal{H}_n$ through parameters in the distribution stage. More details are shown in Materials and Methods.
Then we can bound the  probability that the attacker correctly guesses $\mathbf{X}$ when using an optimal strategy  according to the definition of min-entropy~\cite{konig2009operational},
\begin{equation}
	P_{\rm{guess}}(\mathbf{X}|\mathbf{B})=  2^{-H_{\rm{min}}(\mathbf{X}|\mathbf{B})_\rho}=2^{-\mathcal{H}_n}.
\end{equation}
Thereafter, we can obtain the failure probability of an authentication scenario where Client is the attacker attempting to forge the information sent from Merchant to TP, which is equivalent to the fogery bound in our scheme.~\cite{li2023one}
\begin{equation}
	\epsilon_{ \rm{for}}=m\cdot 2^{1-\mathcal{H}_n}.
\end{equation}

\bigskip
The number of malicious parties is no more than one, i.e., at most one of the above cases happens. Thus, the total security bound, i.e., the maximum failure probability of the protocol, is $\epsilon_{\rm{tot}}= \max\{\epsilon_{\rm{rob}},\epsilon_{\rm{rep}},\epsilon_{\rm{for}}\}$.

\textcolor{black}{From the analysis above, it is obvious that $\epsilon_{\rm{rob}}$ and $\epsilon_{\rm{rep}}$ are constant, while $\epsilon_{\rm{for}}$ is determined by the parameter $n$. In a practical implementation, the users will select a suitable value for $n$ so that $\epsilon_{\rm{tot}}$ satisfies the security requirement. The signature rate of the protocol is $SR=n_x/3n$, where $n_x$ is the total counts under the X basis. We remark that in the analysis we assume that the delay of classical communication is  
negligible and only consider the quantum rate.}

\section{Experimental demonstration}
\begin{figure*}[t]text
	
	\includegraphics[width=\textwidth]{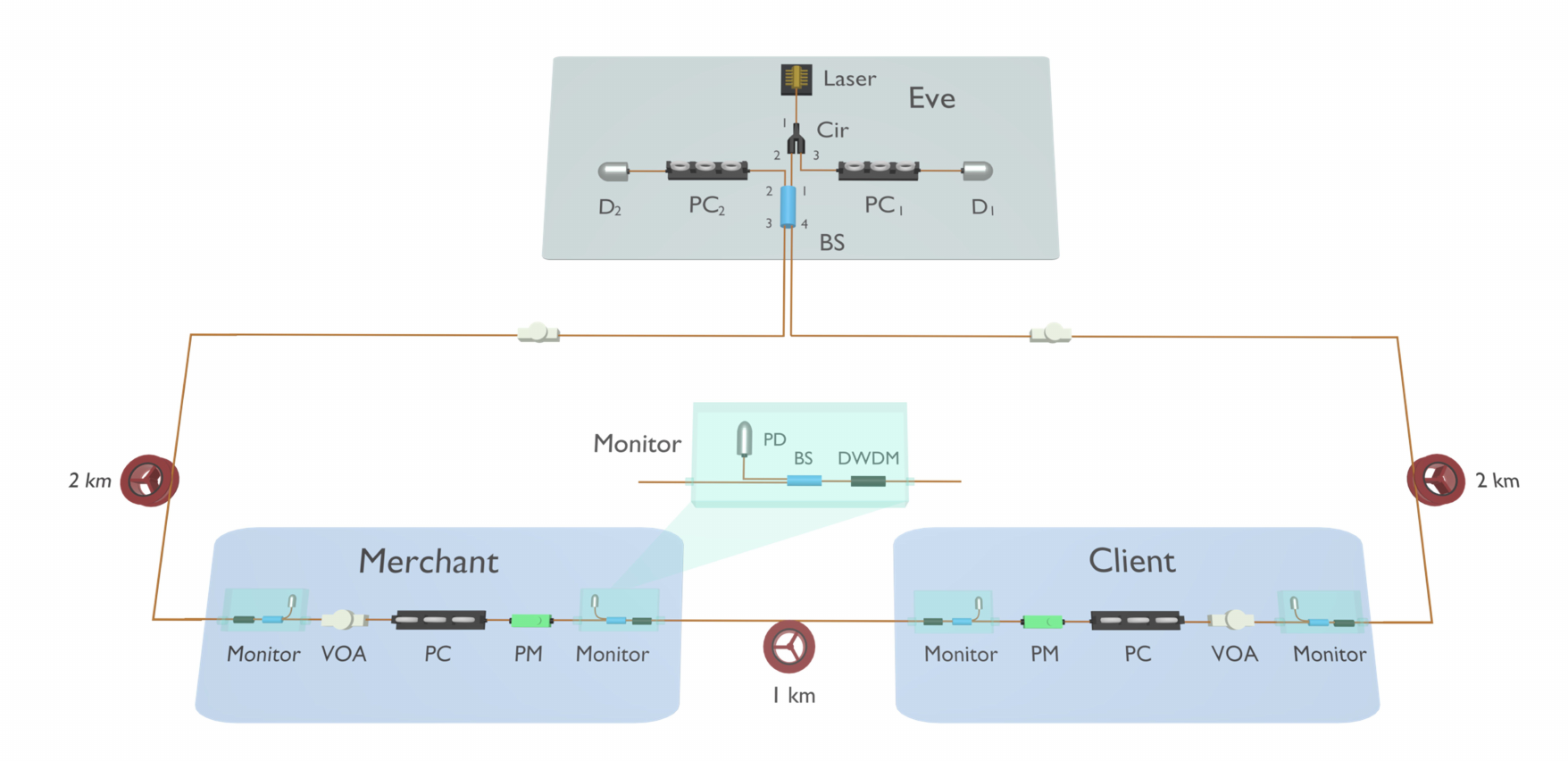}
	\caption{\textbf{Experimental setup of KGP 
     between Merchant and} \boldmath $\rm Client_2$. \textcolor{black}{Here, we take KGP between Merchant and $\rm Client_2$ as an example.} The pulses are generated by a pulsed laser with an extincation ratio of over 30 dB and then split into two pulse sequences with a 50: 50 beam splitter (BS). The pulses entering the loop are subjected to modulation by the phase modulator (PM) operated by either Merchant or Client. Monitor module consists of a dense wavelength division multiplexing (DWDM), a BS and a photon detector (PD). After phase modulation, these two pulses interfere in the Eve's BS and are detected by two superconducting nanowire single-photon detectors $\rm D_1$ and $\rm D_2$. \textcolor{black}{Both the connections between Merchant and Eve and Client and Eve involve 2 kilometers of optical fiber and the total attenuation is 20 dB, achieved through VOA. 
     Besides, there is an insertion of 1 kilometer of optical fiber between Merchant and $\rm Client_2$. VOA: variable optical attenuator; Cir: circulator.} 
	}
	\label{exper}
\end{figure*}

We provide a proof-of-principle demonstration of the entire process here. The scenario considered involves transactional activities between Merchant and Client, with the requirement of a TP to facilitate the purchase transaction. The network consists of two TPs, denoted as $\rm{TP_1}$ and $\rm{TP_2}$ and two clients, denoted as $\rm{Client_1}$ and $\rm{Client_2}$. As illustrated in Fig.~\ref{f1} (C), the channel loss between the Merchant and Client is a constant value of 20 dB, whereas the channel loss between the Merchant and $\rm{TP_1}$ is 15 dB, and that between the Merchant and $\rm{TP_2}$ is 25 dB. \textcolor{black}{Note that We added single-mode optical fiber spools (G.652.D) between the Merchant and $\rm{Client_2}$ (shown in Fig.~\ref{exper}). Specifically, in the Sagnac loop, the lengths of fiber spools are as follows: 2072 meters (0.73 dB) between Merchant and EVE, 2013 meters (0.67 dB) between $\rm{Client_2}$ and EVE, and 1064 meters (0.46 dB) between Merchant and $\rm{Client_2}$. Additional channel attenuation is introduced through variable optical attenuators (VOA).} As there are no assumptions on TP, in practical networks there can be multiple TPs present.

The signature process requires key generation and distribution between Merchant-Client and Merchant-TP, which is facilitated by KGP. In this work, we consider the KGP based on four-phase MDI-QKD~\cite{GU2022}, which remains robust against imperfect devices. 
The system's source flaws are characterized before KGP, including optical power fluctuation, extinction ratio of polarization, phase shift, and pattern effects, and are quantified through a detailed measurement process outlined in Supplementary Materials. The keys used for signature are generated in a plug-and-play system~\cite{PhysRevLett.123.100506,GU2022}, and the global phase of pulses is stabilized by a Sagnac loop. We take KGP between Merchant and $\rm Client_2$ as an example. \textcolor{black}{Note that the experimental setups for KGP of other groups are similar to this one, with the difference being the removal of the optical fibers and polarization controllers (PC) in participants' sites.}

As depicted in Fig.~\ref{exper}, optical pulses modulated by users are generated by an untrusted third party, Eve, and the pulses are separated into two identical pulses.
Merchant ($\rm Client_2$) modulates the clockwise (anticlockwise) pulses. The probability of selecting X basis $p_x$ is 90\% and that of Y basis $p_y$ 
is 10\%. They encode the pulses according to the values of logic bits. After appropriate attenuation, quantum states $\ket{e^{\rm i\theta}\alpha}$ are successfully generated, where $\theta$ is the phase modulated by two participants and the intensity of the pulse is $|\alpha|^2$.
The use of the Sagnac loop has helped to solve the problem of  phase locking, but it has also introduced security concerns due to the possibility of pulses generated by a third party. Therefore, a monitoring module has been added to the participant side to filter and monitor the intensity of the incident pulses, aiming to enhance the security of the system. 
Due to resource limitations, we did not include dense wavelength division multiplexings in the system, and instead opted to replace the photon detectors with power meters during implementation. We would like to emphasize that these modifications have little impact on the results.

Two VOAs are placed between Merchant and Eve, and between $\rm Client_2$ and Eve, respectively, in order to simulate the additional attenuation caused by the communication channels. After interference at Eve's BS, the results are detected by $\rm D_1$ and $\rm D_2$. The time window is selected based on detection data, whose length is 2 ns. The detection efficiency $\eta_{d_1}$ of $\rm D_1$ is $84.4\%$ and the dark count rate $p_{d_1}$ is 4.4 Hz. For $\rm D_2$, the detection efficiency $\eta_{d_2}$ is $85.5\%$ and the dark count rate $p_{d_2}$ is 2.5 Hz. After error correction, Merchant gets $\bf{k_1^{\rm{M}}}$ and Client gets $\bf{k_1^{\rm{C}}}$. The same is for Merchant and TP, where Merchant gets $\bf{k_2^{\rm{M}}}$ and TP gets $\bf{k_2^{\rm{T}}}$.

For demonstration, we sign a file with a size of 0.428 Mb, which is approximately the size of Amazon Web Services Customer Agreement (428072 bits)~\cite{awsurl}.
Merchant generates the signature $\bf{Sig}$ by $Hash(\bf{C}, \bf{k_1^{\rm{M}}}\oplus \bf{k_2^{\rm{M}}})$ and sends it and contract to Client. If Client agrees with the contract, she (or he) will forward $\{\bf{Sig}, \bf{k_1^{\rm{C}}}\}$ to TP. LFSR-based Toeplitz function is used and a detailed description can be seen in Materials and Methods. Upon receiving the signature and contract, TP sends his key $\bf{k_2^{\rm{T}}}$ to Client. Then, Client and TP verify the signature independently by comparing the result of $Hash(\bf{C},\bf{k_1^{\rm{C}}}\oplus \bf{k_2^{\rm{T}}})$ to $\bf{Sig}$. If the signature is successfully verified by both parties, Client will send money to TP, who will in turn pay the Merchant. Otherwise, the contract will be aborted.

\begin{table} 
\renewcommand\arraystretch{1.5}
	\centering
	\caption{\textbf{Four parameters related to the imperfection of realistic sources.} The parameters included the optical power fluctuation $\xi$ , phase shift $\delta$, the extinction ratio of polarization $\tan \theta$, pattern effect $\psi$ and THAs $\mu$. Note that THAs cannot be quantified during our implementation and we set it to a typical value $10^{-7}$~\cite{Lucamarini2015BoundTHA}. }\label{tab03}
    \begin{tabular}{c @{\hspace{0.45cm}} c @{\hspace{0.45cm}} c@{\hspace{0.45cm}} c@{\hspace{0.45cm}} c }\hline \hline
	~&	$\xi$ & $\delta$ & $\tan \theta$  & $\psi$ \\ \hline 
        Merchant - $\rm TP_1$& $0.76\%$& 0.038 & $10^{-2.92}$ & $5.58\times10^{-3}$ \\
        \hline Merchant - $\rm Client_1$& $0.72\%$& 0.035 & $10^{-3}$  & $5.89\times10^{-3}$ \\
        \hline Merchant - $\rm TP_2$& $0.62\%$& 0.035 & $10^{-2.98}$ & $6.91\times10^{-3}$ \\
        \hline \textcolor{black}{Merchant - $\rm Client_2$} & \textcolor{black}{$0.65\%$}& \textcolor{black}{0.037} & \textcolor{black}{$10^{-3.07}$}  & \textcolor{black}{$7.35\times10^{-3}$} \\
		\hline \hline
	\end{tabular}
\end{table}

\section{Experimental results}

Before KGP, we need to characterize the relative parameters corresponding to source flaws of this system. Similar to Ref.~\cite{GU2022}, we denote optical power fluctuation, phase shift, extinction ratio of polarization, and pattern effect $\xi$, $\delta$, $\tan\theta$ and $\psi$, respectively.
Note that although trojan horse attacks (THA) can be resisted in the schemes of two independent users, it cannot be resisted in the plug-and-play system and we set the parameter of THAs to $\mu = 10^{-7}$~\cite{Lucamarini2015BoundTHA}. Results are shown in Table~\ref{tab03}.

\begin{figure*}[t]
\centering
\includegraphics[width=\textwidth]{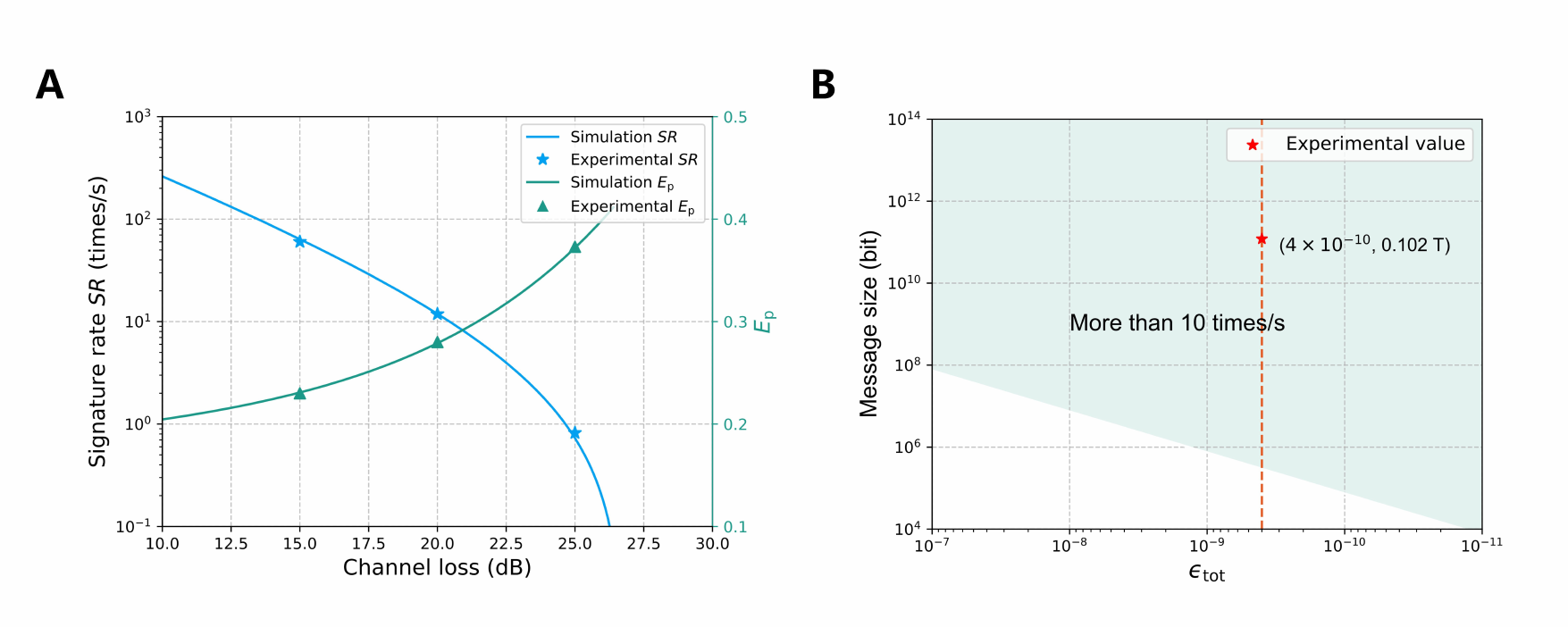}
	\caption{\textbf{Results of demonstration. (A)} Signature rate $R$ under different losses. The total number of pulses sent is $10^{10}$. \textbf{(B)} The relationship between security level and different sizes of files with a key of the same length. The boundary line represents the scenario where the generation rate of keys per second is sufficient to sign a 0.428-Mb file at a security level of $5\times 10^{-10}$ ten times with the same error rate under a 20dB attenuation. In our implementation, a 0.428-Mb document can be signed 11.83 times while maintaining the $5\times 10^{-10}$ security level. Furthermore, with the keys generated within one second, a document of 0.102-terabit (Tb) can be signed ten times at a security level of $4\times 10^{-10}$ ($\epsilon_{\rm rob} = 4\times 10^{-10}$). 
}
	\label{exp_res}
\end{figure*}

The channel loss between  Merchant and Client is fixed at 20 dB. The channel loss between Merchant and TP is set to two scenarios: 15 dB and 25 dB. 
During KGP, the system frequency is 100 MHz and for each pair of participants, the system operates for 100 seconds. Signature rate under different losses (without fiber spools) is shown in Fig.~\ref{exp_res} (A). As a proof-of-principle demonstration, we conducted KGP among different participants within a system. Sagnac loop in the system is used for stabilizing the global phase between two participants and the bit error rate can be kept around 0.10\%. \textcolor{black}{In the scenario where the total length of optical fibers within the loop is approximately 5 km, the bit error rate can still be maintained at around 0.70\%.} Detailed data is shown in Table~\ref{result_table}. Due to the consideration of imperfect sources, the increasing rate of phase error rate is higher compared to protocols of the same type.
\textcolor{black}{For $\rm Client_1$, when signing a 0.428-Mb size document, if $\rm{TP_1}$ is chosen as the third party, the key formed within 100 seconds can be used for signing 1183 times while if $\rm{TP_2}$ is chosen, the number of signature times is 82. Due to the lower signature rate in the case of selecting $\rm{TP_2}$ (25dB), for $\rm Client_2$, the number of signature times remains at 82 when choosing $\rm{TP_2}$.} 
As shown in Fig.~\ref{exp_res} (B), for the same key generation rate, a higher security level implies a smaller signing file size.
There exists a tradeoff between the security level and the size of the signed file. Besides, a minor increase in the signature key length can result in a substantial increase in the size of the signed files, which demonstrates its ability to sign multi-bit files.
The signature rates of the three-party system are limited by the minimum key rate of KGP among the three parties. In practical scenarios, participants can store the keys in advance to reduce the time required for signature generation.

\begin{table*}[t]
			\center
			\caption{\textbf{Summary of experimental data.} We tested the signature rate between different users. The number of total pulses sent is $N = 10^{10}$. The intensity of pulses $\mu$, the experimental bit error rate under X basis and Y basis $E_{\rm b}^x$ and $E_{\rm b}^y$, the total number of leaked bits of information during error correction ${\rm leak_{EC}}$ and signature rate $SR$ are included in the table.} 
   			\setlength{\tabcolsep}{0.3cm} 
      \renewcommand\arraystretch{1.5}
			\begin{tabular}
				{ccccccc}  \hline \hline
				Participants  & $\mu$ & $E_{\rm b}^x$ & $E_{\rm b}^y$&$E_{\rm p}$& ${\rm leak_{EC}}$&  $SR$ \\ \hline
				Merchant - $\rm {TP_1}$ & 7.40$\times 10^{-3}$ &0.10\% &0.07\% &23.0\% & 185875  & 60.10 \\ 
				Merchant - $\rm {Client_1}$  &4.20$\times 10^{-3}$ & 0.10\%&0.06\%&28.0\%  & 59209  & 11.83\\
                Merchant - $\rm {TP_2}$  & 2.30$\times 10^{-3}$ &0.10\%&0.10\%&37.3\%  & 18129  & 0.82\\
                \textcolor{black}{Merchant - $\rm {Client_2}$}  &\textcolor{black}{2.40$\times 10^{-3}$} & \textcolor{black}{0.69\%} &\textcolor{black}{0.55\%}&\textcolor{black}{25.6\%}  & \textcolor{black}{168844}  & \textcolor{black}{4.47}\\
                            \hline \hline
			\end{tabular}
			
			\label{result_table}
\end{table*}

\section{Discussion}

We demonstrate the whole process of quantum e-commerce, which guarantees the one-time of purchase with information-theoretic security. As the length of the signature increases, the probability of cheating approaches zero. The QDS scheme can complete multiple contract signatures within one second, further narrowing the gap between quantum and classical schemes.

We have considered a real-world scenario where Merchant and Client need to carry out a transaction and establish a consensus to complete it. TP is introduced to finish the process. Unlike the classical setup where third parties are usually considered trustworthy by default, there are no assumptions on TP. We take the Amazon Web Services Customer Agreement~\cite{awsurl} with the size of 428072 bits and demonstrate the underlying principles of the process.
Note that the size of the contract affects the final performance instead of the content. The signature rates of the system are limited by the minimum key rate
of KGP among the three parties. Consequently, for $\rm Client_1$, the signature rate of a 0.428-Mb document can reach 11.83 times per second when choosing $\rm TP_1$ and 0.82 times per second with $\rm TP_2$ \textcolor{black}{and for $\rm Client_2$, the signature rate can reach 4.47 times per second when choosing $\rm TP_1$ and 0.82 times per second with $\rm TP_2$.}

The proposed quantum e-commerce scheme, employing QDS, offers a solution for ensuring message authenticity and integrity in the presence of imperfect keys and devices. The elimination of privacy amplification reduces the computational resources and running time of postprocessing. Furthermore,a thorough experimental characterization of source flaws in the scheme is conducted. This feature distinguishes our scheme as a practical solution for addressing the issue of imperfect keys and devices in the field of quantum communication.
The proposed scheme also demonstrates robustness against security levels and finite-size effects, making it highly compatible with future quantum networks and suitable for a variety of applications.

In summary, the proposed scheme can accomplish the task 
of e-commerce with practical devices and outperforms other quantum protocols. We have validated the effectiveness of our scheme through the demonstration of an e-commerce scenario, involving a transaction between Merchant and Client that required a consensus to complete. It presents a promising approach to ensuring message authenticity and integrity in the presence of imperfect keys and devices.

\section*{Acknowledgments}

This study was supported by the National Natural Science Foundation of China (No. 12274223), the Natural Science Foundation of Jiangsu Province (No. BK20211145), the Fundamental Research Funds for the Central Universities (No. 020414380182), and the Program for Innovative Talents and Entrepreneurs in Jiangsu (No. JSSCRC2021484).

\appendix

\section{Four-phase MDI-KGP between Merchant and Client}\label{ap:KGP}
During implementation, Merchant, Client, and TP send their quantum states to the intermediates who perform interference measurements on the received pulses and announce the outcomes. These three parties will later distill their keys through their quantum states and measurement results announced by the intermediates. This process, in detail, is equivalent to four-phase MDI-KGP between Merchant and Client, and between Merchant and TP. 

Here, we give a brief introduction to four-phase MDI-KGP in Merchant-Client channel to show the details of steps (i)-(iii) in protocol description. 

(i) \emph{Preparation.} Both of Merchant and Client randomly choose the $\rm X$ and $\rm Y$ bases with probabilities $p_{x}$ ($0< p_x < 1$) and $p_{y}=1-p_{x}$, respectively. For $\rm X$ basis, Merchant encodes a coherent state $\ket{e^{{\rm i}k_{x}^{\rm M}\pi}\alpha}$ with the logic bit $k_{x}^{\rm M}\in\{0,1\}$, where $|\alpha|^2$ is the intensity of the optical pulse. For $\rm Y$ basis, Merchant prepares a coherent state $\ket{e^{{\rm i}(k_{y}^{\rm M}+\frac{1}{2})\pi}\alpha}$ according to random logic bit $k_{y}^{\rm M}\in\{0,1\}$.
Client prepares his own state $\ket{e^{{\rm i}k_{x}^{\rm C}\pi}\alpha}$ or  $\ket{e^{{\rm i}(k_{y}^{\rm C}+\frac{1}{2})\pi}\alpha}$ according to the same rule. 
Then, Merchant and Client send their optical pulses to an untrusted relay, Eve, through insecure quantum channels.

(ii) \emph{Measurement.}  Eve performs interference measurements on the received pulses with a 50: 50 BS and two single-photon detectors, denoted as D$_1$ and D$_2$, and records the detection results. Those where one and only one detector clicks are defined as an effective measurement. 

(iii) \emph{Sifting.} Merchant and Client repeat Steps (i)–(ii) for N times. $\ket{\bf{X_1}}=\otimes^{N}_{i=1}\ket{i},~\ket{i}\in\{\ket{e^{{\rm i}k_{x}^{\rm M}\pi}\alpha},~\ket{e^{{\rm i}(k_{y}^{\rm M}+\frac{1}{2})\pi}\alpha}\}$ is all Merchant's states in direct product state. Likewise, $\ket{\bf{Y_1}}=\otimes^{N}_{j=1}\ket{j},~\ket{j}\in\{\ket{e^{{\rm i}k_{x}^{\rm C}\pi}\alpha},~\ket{e^{{\rm i}(k_{y}^{\rm C}+\frac{1}{2})\pi}\alpha}\}$ is all Client's states in direct product state. 
Eve announces the location of all effective measurements and which detector (D$_1$ or D$_2$) clicks. For every effective measurement announced by Eve, if D2 clicks, Client will flip his corresponding logic bit ($k_{x}^{\rm C}$ or $k_{y}^{\rm C}$).
Merchant and Client will only keep their logical bits of effective measurements and discard other bits.
Then they disclose their basis choices for effective measurements through authenticated classical channels, and further classify their key bits with basis information.

(iv) \emph{Parameter estimation.}
Merchant and Client publicize all their bits in the Y basis to calculate the bit error rate $E^y_{\rm b}$, and also obtain the number of counts $n_x$ and $n_y$ under X and Y bases, respectively.

(v) \emph{Key distillation.} Merchant and Client perform error correction on the remaining keys under the X basis with $\varepsilon_{\rm{cor}}$-correctness to obtain the final keys.

Merchant then randomly disturbs the orders of his final key string, and publicizes the new order to Client through authenticated channels.
Subsequently, Client change the orders of his key strings according to the order announced by Merchant.
Finally, Merchant and Client divide their final keys into $3n$-bit strings, each of which is used to sign a message.
Each of $\bf{k_1^{\rm{M}}}$, $\bf{k_2^{\rm{M}}}$, $\bf{k_1^{\rm{C}}}$, and $\bf{k_2^{\rm{T}}}$ is one string.

\section{LFSR-based Toeplitz hash functions}\label{hash}
In the protocol, we utilize universal hash functions to generate the signature. Concretely, we choose 	LFSR-based Toeplitz hash function. In the description of the proposed scheme, we combine the process of XOR encryption on hash value into the hash function and express it as $Hash(\bf{C},\bf{k})$ for simplicity.
To show details of this function, we rewrite is as
$Hash(\bf{C},\bf{k})$$=Hash(\bf{x_1}, \bf{x_2}, \bf{x_3}, \bf{x_4})$, where $\bf{C}=\bf{x_1}$$\in\{0,1\}^{|\bf{C}|}$ corresponds to the contract and $\bf{k}=(\bf{x_2}, \bf{x_3}, \bf{x_4})$$\in\{0,1\}^{3n}$ corresponds to keys in step (ii) of protocol. The lengths of $\bf{x_2}, \bf{x_3}, \bf{x_4})$ are all n-bit.
Denote the length of $\bf{x_1}$ as $m$, i.e., $|\bf{C}|=m$. Then
\begin{equation}
Hash(\bf{x_1}, \bf{x_2}, \bf{x_3}, \bf{x_4})=\bf{H_{nm}}\cdot \bf{x_1} \oplus \bf{x_4}, 
\end{equation}
where $\bf{H_{nm}}$$=f(\bf{x_2}, \bf{x_3})$ is a LFSR-based Toeplitz matrix determined by $\bf{x_2}$ and $\bf{x_3}$. The random string $\bf{x_2}$ maps a random irreducible polynomial $p(x)=x^n+p_{n-1}x^{n-1}+...+p_1x+p_0$ in GF(2) of order $n$ that decide the structure of LFSR. Details of generating a random irreducible polynomial can be found in Supplementary Materials. Another random string $\bf{x_3}=(a_n,a_{n-1},...,a_2,a_1)^T,~ a_i\in\{0,1\}$ is the initial state. LFSR will expand the initial state into a matrix $\bf{H_{nm}}$ with $n$ rows and $m$ columns. The structure of LFSR can be represented as an $n\times n$ matrix
\begin{equation}
	\bf{W}=\begin{pmatrix}
		p_{n-1}& p_{n-2} & ... & p_1 &p_{0} \\
		1 & 0 & ... & 0 & 0 \\
		0 & 1 & ... & 0 & 0 \\
		...&...&...&...&... \\
		0 & 0 & ... & 1 & 0
	\end{pmatrix}.
\end{equation}

The structure of LFSR-based Toeplitz matrix can be represented through $\bf W$ as $\bf{H_{nm}}$$=\left(\bf{x_3},\bf{W x_3},...,\bf{W}^{m-1} \bf{x_3}\right)$
~\cite{krawczyk1994lfsr}.

\section{Calculation details}\label{calcul}
Here, we give an introduction to the calculation details of KGP used in the scheme. 
According the Ref.~\cite{GU2022},
only key bits under the X basis are used to form secure key bits. Since the privacy amplification step is removed in our scheme, the total unknown information of the $l$-bit string such as $x_i,~i\in \{2,3,4\}$ in LFSR-based Toeplitz hash functions considering finite-key effect is given by
\begin{equation}\label{finite}
	\begin{aligned}
	    \mathcal{H}_n=l\cdot \Bigg[1-H\left(\overline{E}^l_{\rm p}\right)-\frac{1}{n_x}{\rm leak}_{\rm EC}-\frac{1}{n_x}{\rm log}_2\frac{2}{\epsilon_{\rm EC}}\Bigg],
	\end{aligned}
\end{equation}
where $n_{x}$ is the total counts under the X basis. $\overline{E}^n_{\rm p}$ is the upper bound of the phase error rate of all counts in an $n$-bit string. ${\rm leak}_{\rm EC}=n_xfH(E_{\rm b}^{x})$ is the number of leaked bits of information during the error correction, where $f$ is the error correction efficiency and $E_{\rm b}^{x}$ is the bit error rate in X basis. $H(x)=-x\log_{2}x-(1-x)\log_2(1-x)$ is the binary Shannon entropy function. $\overline{E}^l_{\rm p}$ can be given by

\begin{equation}
    \overline{E}^l_{\rm p}=\overline{E}_{\rm p} +	\gamma^U \left(l, n_x-n, E_{\rm p}, \epsilon \right),
\end{equation} 
where $	\gamma^{U}(l,k,\lambda,\epsilon)=\frac{\frac{(1-2\lambda)AG}{l+k}+
\sqrt{\frac{A^2G^2}{(l+k)^2}+4\lambda(1-\lambda)G}}{2+2\frac{A^2G}{(l+k)^2}}$ represents the statistical fluctuation of the random sampling without replacement, with $A=\max\{l,k\}$ and $G=\frac{l+k}{lk}\ln{\frac{l+k}{2\pi lk\lambda(1-\lambda)\epsilon^{2}}}$~\cite{yin2020tight}, and $\overline{E}_{\rm p}$ is the upper bound of phase error rate.
$E_{\rm p}$ satisfies the following inequality~\cite{sun2021security,GU2022}
\begin{equation}\label{epbound}
	1-2\Delta\leq\sqrt{E_{\rm b}^{y}E_{\rm p}}+\sqrt{(1-E_{\rm b}^{y})(1-E_{\rm p})},
\end{equation}
where $\Delta$ quantifies the imbalance of Alice's and Bob's quantum coins according to their basis selection. $E_{\rm b}^{y}$ represents the bit error rate in the $\rm Y$ basis.
In the symmetric scenario, the relation between $\Delta$ and fidelity can be simplified as~\cite{tamaki2012erratum} $$1-2Q\Delta={\rm Max}_{\delta_{\theta}, \delta_{\rm X}, \delta_{\rm Y}}{\rm Re}\left(e^{i\delta_\theta}\fid{\Psi_{{\rm Y}, \delta_{\rm Y}}}{\Psi_{{\rm X}, \delta_{\rm X}}}\right)|\fid{\Psi_{\rm Y}}{\Psi_{\rm X}}|,$$ 
where $Q$ is the gain and $\delta_{\theta}$, $\delta_{\rm X}$, and $\delta_{\rm Y}$ are free variables, the values of which range from 0 to $2\pi$. Besides, $\Ket{\Psi_{{\rm Z}, \delta_{\rm Z}}}$ is $(\Ket{0_{\rm Z}}\Ket{\Psi_{0\rm Z}}+\Ket{1_{\rm Z}}\Ket{\Psi_{1\rm Z}})/\sqrt{2}$. $\Ket{\Psi_{i\rm Z}}$ means the state is prepared under the Z basis (Z$\in\{\rm{X, Y}\}$) with the bit value $i$ ($i \in\{\rm{1, 0}\}$).
The fidelity with imperfect sources can be expressed as 
\begin{equation}\label{practicalfinal}
	\begin{aligned}
		\fid{\Psi_{\rm X}}{\Psi_{\rm Y}}=&\frac{1}{4}(1-\epsilon)e^{-\mu}{\rm cos}^2\theta \bigl[(1-\rm i)\fid{\alpha^\prime}{ie^{i\delta}\alpha^\prime}\\
		&+(1- \rm i)\fid{-e^{i\delta}\alpha^{\prime}}{-ie^{i\delta}\alpha^{\prime}}\\
		&+(1+ \rm i)\fid{\alpha^{\prime}}{-ie^{i\delta}\alpha^{\prime}}\\
		&+(1+ \rm i)\fid{-e^{i\delta}\alpha^{\prime}}{ie^{i\delta}\alpha^{\prime}}\bigr],
	\end{aligned}
\end{equation}
where $\epsilon$, $\mu$, $\theta$, and $\delta$ represent pulse correlations, THAs, side channels in the polarization space, and phase shift, respectively. Note that $\alpha^{\prime}$ is the actual intensity of pulses. According to the formulas above, we can get $\Delta$. 

Considering the finite key effect, we use the bound of the concentration inequality~\cite{kato2020concentration, GU2022} to derive the upper bound of $E_{\rm p}$. The inequality is 
\begin{equation}\label{S10}
		\Lambda_n\leq \sum_{u=1}^{n}{\rm Pr}(\xi_u=1|\xi_1,..., \xi_{u-1})+\Delta_n,
\end{equation}
where $\xi_1,..., \xi_n$ is a sequence of Bernoulli random variables and $\Lambda_n=\sum_{u=1}^{n} \xi_u$. $\Delta_n=\sqrt{\frac{1}{2}n{\rm ln}\epsilon_{\rm F}^{-1}}$, where $\epsilon_{\rm F}$ is the failure probability.
$m_y = n_y E_b^y$ is the number of bit errors in the Y basis and the upper bound of the expectation value  $m_y^{*}$ is $m_y + \Delta_{n_y}$. Thus, ${E_{\rm b}^{y}}^{*}=m_y^{*}/n_y$ and then $E_{\rm p}^{*}$ can be calculated according to Eq.~\ref{epbound}. So $m_{\rm p}^{*}=n_x E_{\rm p}^{*}$. Then estimate the upper bound $\overline{m}_x$ through the concentration inequality. $\overline{E}_{\rm p}$ can be naturally calculated with $\overline{E}_{\rm p}=\overline{m}_{\rm p}/n_x$. We set all failure probabilities $\epsilon_{\rm EC}=\bar{\epsilon}=\epsilon_{\rm F}=10^{-10}$ during calculation. Parameter $n$ is optimized with the constraint that $\epsilon_{\rm{tot}}\le 5\times10^{-10}$.

\section{Error correction algorithm}
In our implementation, the error correction algorithm is used to ensure that the keys generated by both parties through the KGP process are completely identical and the algorithm we use is Cascade algorithm~\cite{brassard1994secret}. The block size for each error correction is set to 1M and the size of the remaining keys is smaller than 1M and these keys are corrected together. The detailed process is presented as follows:

(i) Alice and Bob randomly permute the original keys based on a pre-agreed random sequence and record the permutation information.

(ii) The permuted keys are divided into different segments of a fixed length. We set the length 600.

(iii) The parity check codes for each segment are computed, and both parties compare them over a publicly authenticated channel.

(iv) For segments with consistent parity check codes, no further processing is performed. For segments with different codes, error correction is performed using binary search.

(v) When the iteration number is greater than 1, based on the recorded random permutation information from the previous round, the position of the key belonging to the error bit can be identified. Since in the previous round, all segments have the same parity check code, indicating either no errors or an even number of errors, if an error bit is discovered in this round, another error bit can be found. Another error bit is found using a binary search algorithm. This process continues until no further error bits can be found.

(vi) The above steps are repeated until the parity check codes for all segments are completely identical.

During implementation, we do not aim for optimal error correction performance, and the number of iterations is limited to a maximum of three. The error correction efficiency $f$ is not more than 1.13 during our implementation.

%


\clearpage

\section*{Supplementary Materials}

\section*{section S1. Irreducible polynomials}
Irreducible polynomials play an important role in hashing process. The linear feedback shift register (LFSR) Toeplitz hash function is decided by a randomly selected irreducible polynomial of order $n$ in GF(2) and an initial random bit string. 
Here we introduce the definition and criteria of irreducible polynomials, and solutions to randomly generating an irreducible polynomials in GF(2).\\
\emph{A. Introduction to irreducible polynomials in GF(2)}\\
All polynomials in GF(2), with coefficients only being `0' or `1', composite a ring, where all calculation obeys the rules in GF(2). A polynomial in GF(2) is irreducible means that no polynomials can divide it except the identity element `1' and itself, under calculation rules in GF(2).
Denote a polynomial of order n in GF(2) as $p(x)$. There is another necessary and sufficient condition for $p(x)$ being irreducible that
\begin{equation}\label{p1}\tag{S1}
	\left\{
	\begin{gathered}
		x^{2^n} \equiv x ~~\rm{mod} ~p(x)
		\\[3pt]
		\rm{gcf}(x^{2^{\frac{n}{d}}}-x,p(x))=1
	\end{gathered},
	\right.
\end{equation}
for all $d$ that is prime factor of $n$, where $\rm{gcf} ( f(x),g(x) )$ represents the greatest common factor of $f(x)$  and $g(x)$.\\
\emph{B. Generating an irreducible polynomials in GF(2) in random}\\
The condition above offers a way to testing the irreducibility of a polynomial. Trivially, there is a solution to randomly generating an irreducible polynomials in GF(2). One can directly generate a polynomial through random numbers and test for its irreducibility by repeating this step until a polynomial passes the test. We denote this solution as the test algorithm. When $n=2^k$, parameter $d$ can only be 2, and thus the test algorithm is quite efficient by utilizing fast modular composition algorithm and extended Euclidean algorithm~\cite{yinnwac228}. However, under other scenarios the test may consume too much computational resources and also require too many random numbers. In the following we introduce another approach with higher efficiency~\cite{shoup1996fast}, denoted as the generating algorithm.
This solution requires presetting an irreducible polynomial of order $n$, defining the extension field GF($2^n$). Another random number decides a random element in GF($2^n$). The minimal polynomial of this random element in this field GF($2^n$) must be irreducible. Thus, by computing the minimal polynomial of this element one can get a random irreducible polynomial.  An efficient process to obtain the minimal polynomial is shown below.

First, denote the initial irreducible polynomial as $f(x)$ and the polynomial generated by random element as $g(x)$. Then calculate the sequence $a_0=g_0(0)$, $a_1=g_1(0)$, ..., $a_{2n-1}=g_{2n-1}(0)$ in turn, where $g_i(x)=g^i(x)$ mod $f(x)$. This sequence of $2n$ elements can fully determine the minimal polynomial of $g(x)$, i.e., the monic polynomial $h(x)$ of least order such that $h(g(x))=0$ mod $f(x)$, which can be efficiently computed by Berlekamp-Massey algorithm \cite{massey1969shift}. The result, i.e., the minimal polynomial of $g(x)$, will be the random irreducible polynomial we generate. The probability of this polynomial to be $n$-order is more than $1-2^{-n/2}$. If the order is less than $n$, we can just choose another random number and repeat the process.

Here, we give a demonstration of the generating algorithm for order $n=8$ as an example. 
First, preset the initial irreducible polynomial $f(x)=x^8+x^7+x^6+x+1$, 
and input a random bit string $(0111 1100)$, corresponding to $g(x)=x^6+x^5+x^4+x^3+x^2$.
Then calculate $g_2(x)=g^2(x)$ mod $f(x)=x^6+x^5+x^4+x^3+x^2$, $g_3(x)=g^3(x)$ mod $f(x)=x^4+x$, ..., $g_{15}(x)=g^{15}(x)$ mod $f(x)=x^7+x^6+x^5+x^3+x^2+x$. Then we can obtain a sequence $\bf{s}$ with $\bf{s}$$(0)=g_0(0)=1$, $\bf{s}$$(1)=g_1(0)=0$, $\bf{s}$$(2)=g_2(0)=0$, ..., $\bf{s}$$(15)=g_{15}(0)=0$, i.e., $\bf{s}$= $(1000 0010 1011 1110)$. Finally, input $\bf{s}$ into Berlekamp-Massey algorithm with GF(2), and we can obtain the output $(101111011)$, corresponding to the generated random irreducible polynomial $h(x)=x^8+x^6+x^5+x^4+x^3+x+1$.

The test algorithm can only be effectively accelerated when the order $n=2^k$~\cite{yinnwac228}. The generating algorithm will be more efficient than the test algorithm even under this circumstance. We simulate the time consumption of the generating algorithm and the test algorithm, with $n=256$ as an example. The results shows that it takes 1016.57 seconds to generate twenty random irreducible polynomials through the test algorithm, and only 133.4 seconds if using the generating algorithm. The time consumption of the test algorithm will be longer if $n\ne 2^k$. It shows that the generating algorithm is quite efficient.

\section*{section S2. Characterization of source flaws}
As a MDI type protocol, our protocol has no assumptions on the middle node, which means that we only need to consider the imperfections at the source. Since KGP in our scheme is based on four-phase MDI QKD, our protocol inherits its advantages and the way to characterize the source flaws. Here, we show how to get the parameters of the source flaws in this section. We consider the worst case scenario here and the structure and devices used in this setup are identical to those used in the plug-and-play system for key generation, with the only difference being the removal of extra attenuation. The quality of the radio frequency modulation signals for the modulation and synchronization of the system has a direct impact on the results of characterization. All radio frequency signals we used were produced
by an arbitrary waveform generator with a sampling rate
of 2.5 $\rm G \cdot \rm Sa$/s (Tabor Electronics, P2588B). Note that random numbers are generated by a home-made quantum random number generator and the cycle length is 10000. The electrical drivers we used are DR-VE-10-MO (iXblue), and the ultra-low voltage PMs are PM-5S5-10-PFA-UV (Eospace).
During our implementation, we conducted four rounds of KGP, and we also performed four rounds of calibrations of the source, whose results showed minor differences.\\
\subsection*{A. State preparation flaws}
State preparation flaws results from imperfect intensity modulation and phase modulation and the pratical state can be written as~\cite{GU2022}:
\begin{equation}\label{statflaw}\tag{S2}
	\begin{aligned}
		\Ket{\Psi^{*}_{\rm X}}=&\frac{1}{\sqrt{2}}\left( \ket{0_{\rm X}}\ket{\alpha_0}+\ket{1_{\rm X}}\ket{-e^{\rm i(\pi+\delta_1)}\alpha_1} \right),\\
		\Ket{\Psi^{*}_{\rm Y}} = &\frac{1}{\sqrt{2}}\left( \ket{1_{\rm Y}}\ket{\rm ie^{i(\pi/2+\delta_2)}\alpha_2}+ \right.
        \left. \ket{0_{\rm Y}}\ket{-{\rm i}e^{\rm i(3\pi/2+\delta_3)}\alpha_3} \right).
	\end{aligned}
\end{equation}
We define $\delta_i$ as the difference between the actual phase and the expected phase, $i \in \{1, 2, 3\}$.
In our scheme, there is no need for multi-intensity modulation and we only need to consider imperfect phase modulations, which is named phase shift. Another point is optical power fluctuation, caused by the volatility of the path attenuation and the light source itself. We show how to quantify specific parameters through experiments in the following.\\
\emph{(i) Optical power fluctuation}\\
During our implementation, the intensity of pulses is a constant under a specific channel loss. Nevertheless, due to the inherent fluctuations of the light source and path variations, the intensity of the light is not a constant.
$|\alpha^\prime|^2$ represents the realistic intensity modulation of coherent pulses. The deviation ratio can be denoted by $\xi^\prime=\left||\alpha^\prime|^2-|\alpha|^2\right|/|\alpha|^2$ and $\xi$ is the maximum value of $\xi^\prime$.
With no intensity modulation, we can quantify this parameter by measuring the optical power fluctuations. We recorded the values of the optical power meter (Joinwit, JW8103D) every 500 ms for 600 seconds.

We separated the optical power fluctuations into two parts: the intrinsic fluctuation of the light source and the path fluctuation. Note that the added attenuation also causes fluctuation and we distributed the attenuation into two measurement processes to ensure that the values of optical power are not only greater than the power meter's minimum detectable value, but also the pulse intensity after the total attenuation is lower than the minimum value used in the experiment. We measured the fluctuation of the light source by directly connecting the light source to a power meter after introducing an additional attenuation of approximately 40 dB, where the attenuation values slightly vary among the four groups. While for path fluctuation, we directly disconnected the optical loop and connected a power meter at the output port of a participant to measure the optical power fluctuations after the pulses passed through the entire Sagnac loop. Different attenuations were also introduced to facilitate their display in Fig.~S1. The maximum values of $\xi$ in different pairs of participants (Merchant - $\rm {TP_1}$, Merchant - $\rm Client_1$, Merchant - $\rm {TP_2}$ and Merchant - $\rm Client_2$) are 0.76\%, 0.72\%, 0.62\% \textcolor{black}{and 0.65\%.}\\
\emph{(ii) Phase shift}\\
From Eq.~\ref{statflaw}, we can know imperfect phase modulation can be quantified by $\delta = \max\left\{\overline{\delta_1}, \overline{\delta_2}, \overline{\delta_3}\right\}$. Similar to Refs.~\cite{PhysRevA.92.032305, GU2022}, the structure and parameters of the measurement system are the same as the plug-and-play system. During implementation, Merchant performs phase modulation randomly while Client sets his PM at a fixed phase 0. Two pulse trains interfere at Eve's BS and the interference results are detected by single-photon detectors. Here, the random number resource used by Merchant is the same as that used in the KGP and the intensity of pulses is set as 0.001. We ran the system for 100 seconds and recorded the corresponding results. 
We employed Hoeffding's inequality~\cite{hoeffding1994probability} to get the upper bound of $\delta_\phi$ (Eq.~\ref{uppb}), where $\eta_{d_i}$ is the detection efficiency of detectors and $\overline{D}_{i,\phi} = D_{i,\phi} + \sqrt{D_{i,\phi}/2\ln(1/\varepsilon)}$ ($i \in \{1, 2\}$) and $\varepsilon = 10^{-10}$ is the failure probability. For $\phi\in [0,\pi]$, we have $\phi_0 = \phi$, and for $\phi \in(\pi,2\pi]$, $\phi_0 = \phi - \pi$. The detailed detection results and their upper bounds of phase shift are presented in Table.~S1. Consequently, the values of $\delta$ are $0.038$, $0.035$, $0.035$ \textcolor{black}{and $0.037$,} respectively. 

\begin{figure*}[htb] 
	\centering
	\begin{equation}
		\overline{\delta}_\phi ={\rm max}\left\{ \left |\phi_0 - 2{\rm arctan}\sqrt{\frac{(\rm \overline{D}_{2,\phi}-\rm \underline{D}_{2,0})/\eta_{d_2}} {(\rm \underline{D}_{1,\phi}-\rm \overline{D}_{2,0})/\eta_{d_1}}}\right |, \left |\phi_0 - 2{\rm arctan}\sqrt{\frac{(\rm \underline{D}_{2,\phi}-\rm \overline{D}_{2,0})/\eta_{d_2}} {(\rm \overline{D}_{1,\phi}-\rm \underline{D}_{2,0})/\eta_{d_1}}}\right | \right\}. \label{uppb}\tag{S3} 
	\end{equation}
\end{figure*}

\subsection*{B. Side channels caused by polarization}
Various kinds of side channels caused by mode dependencies exist in a QKD system, through which the eavesdropper can distinguish the states. In our scheme, we only consider the polarization space as the side-channel space and the coherent states can be written as 
\begin{equation}
	\begin{aligned}
		&\ket{\alpha_0^{\prime}}=\rm cos\theta_0\ket{\alpha_0}_{\rm H}+\rm sin\theta_0\ket{\alpha_0}_{\rm V},\\	
		&\ket{\alpha_1^{\prime}}=\rm cos\theta_1\ket{-e^{i(\pi/2 +\delta_1)}\alpha_1}_{\rm H}+\rm sin\theta_1\ket{-e^{i(\pi/2 + \delta_1)}\alpha_1}_{\rm V},\\
		&\ket{\alpha_2^{\prime}}=\rm cos\theta_2\ket{ie^{i(\pi + \delta_2)}\alpha_2}_{\rm H}+\rm sin\theta_2\ket{ie^{i(\pi + \delta_2)}\alpha_2}_{\rm V},\\
		&\ket{\alpha_3^{\prime}}=\rm cos\theta_3\ket{-ie^{i(3\pi/2+\delta_3)}\alpha_3}_{\rm H}+\rm sin\theta_3\ket{-ie^{i(3\pi/2+\delta_3)}\alpha_3}_{\rm V},\\
	\end{aligned}\tag{S4}
\end{equation}
where H and V represent two orthogonal states. As we consider the worst scenario where the maximum deviation $\theta$ applies to all states here, we can assume $\theta_i = \theta$ ($i\in \{0,1,2,3\}$). Thus, the states with side channels and state preparation flaws can be expressed as
\begin{equation}
	\begin{aligned}
		&\Ket{\Psi^{\prime}_{\rm X}}=\frac{1}{\sqrt{2}}\left( \ket{0_{\rm X}}\ket{\alpha_0^{\prime}}+\ket{1_{\rm X}}\ket{\alpha_1^{\prime}} \right),\\
		&\Ket{\Psi^{\prime}_{\rm Y}}=\frac{1}{\sqrt{2}}\left( \ket{1_{\rm Y}}\ket{\alpha_2^{\prime}}+\ket{0_{\rm Y}}\ket{\alpha_3^{\prime}} \right).\\	
	\end{aligned}\tag{S5}
\end{equation}

For quantification, we only need to get the maximum deviation of $\theta$. 
As we used the horizontal states to form secret key, the extinction ratio of polarization in the system can be represented by $\tan\theta$. 
The extinction ratio of polarization of the sent pulses is the ratio of the optical power along the fast-axis to the slow-axis.
To obtain this ratio, a polarization beam splitter (PBS) is placed at a participant's site and we measured the optical power at both ports of the PBS using a power meter. Note that in the case involving 5-km optical fibers, an additional polarization controller is introduced.
The measurements were taken once every 500 ms and were carried out for 600 seconds. As shown in Fig.~S2, the orange points mean the extinction ratio of polarization.
The maximum values of the extinction ratio of pulse polarization in four rounds of KGP are -29.2 dB, -30 dB, -29.8 dB \textcolor{black}{and -30.7 dB,} and thus the values of $\tan \theta$ are $10^{-2.92}$, $10^{-3}$, $10^{-2.98}$ \textcolor{black}{and $10^{-3.07}$,} respectively.

\subsection*{C. Trojan horse attacks}
Here, we consider a type of THA where an eavesdropper, Eve, injects strong light into the optical devices at the participants' sites and obtains modulation information by measuring the reflected light. To model this type of attack, the pulse state $\ket{\xi}_{\rm E}$ held by Eve can be expressed as $e^{-\mu/2}\ket{\alpha_{\rm Z}}+\sqrt{1-e^{-\mu}}\ket{\beta_{\rm Z}}$, where $T\in \{0_x,0_y,1_x,1_y\}$ represents the bit value and basis selection, and $\mu$ is the intensity of Eve's back-reflected pulses~\cite{pereira2019quantum}. In the worst case scenario, where $\fid{\xi_{\rm T}}{\xi_{{\rm T}^{\prime}}}_{\rm E}=0$, and considering the state preparation flaws and side-channel effects, the states can be expressed as

\begin{equation}
	\begin{aligned}
		&\Ket{\Psi^{\prime\prime}_{\rm X}}=\frac{1}{\sqrt{2}}\left( \ket{0_{\rm X}}\ket{\alpha_0^{\prime}}\Ket{\zeta_{\rm 0X}}_{\rm E}+\ket{1_{\rm X}}\ket{\alpha_1^{\prime}}\Ket{\zeta_{\rm 1X}}_{\rm E} \right),\\
		&\Ket{\Psi^{\prime\prime}_{\rm Y}}=\frac{1}{\sqrt{2}}\left( \ket{1_{\rm Y}}\ket{\alpha_2^{\prime}}\Ket{\zeta_{\rm 1Y}}_{\rm E}+\ket{0_{\rm Y}}\ket{\alpha_3^{\prime}}\Ket{\zeta_{\rm 0Y}}_{\rm E} \right).\\	
	\end{aligned}\tag{S6}
\end{equation}

Note that the scheme with two independent users can resist THAs. However, the plug-and-play system is vulnerable to such attacks, since the pulses are generated by a third party. We set $\mu = 10^{-7}$~\cite{Lucamarini2015BoundTHA} to illustrate the robustness of our scheme to such attacks.

\subsection*{D. Pattern effect}
Security analyses of quantum protocols tend to be based on the assumption of independent distributed pulses. However, pulse correlation resulting from band-limited devices is almost impossible to avoid in practical systems. In this way, as the system frequency increases, the value of $\psi$ will rapidly increase, significantly influencing the signature rate. Encoding logic bits with phase information removes the requirements of vacuum state preparation in our protocol, which makes our scheme free from intensity correlation~\cite{yoshino2018quantum}, and the pulse correlation, also called pattern effect, caused by phase modulation can be observed in our implementation. According to Refs.~\cite{Pereiraeaaz4487,GU2022}, the state can be written as 
\begin{equation}\label{pattstate}
		\begin{aligned}
			&\Ket{\Psi_{\rm X}}=\sqrt{1-\epsilon}\Ket{\Psi^{\prime\prime}_{\rm X}}+\sqrt{\epsilon}\Ket{\Psi_{\rm X}^{\prime\prime\perp}},\\
			&\Ket{\Psi_{\rm Y}}=\sqrt{1-\epsilon}\Ket{\Psi^{\prime\prime}_{\rm Y}}+\sqrt{\epsilon}\ket{\Psi_{\rm Y}^{\prime\prime\perp}},\\
		\end{aligned}\tag{S7}
	\end{equation} 
where $\epsilon$ is the parameter characterizing the pulse correlation and $\ket{\Psi_{\rm Z}^{\prime\prime\perp}}$ ($\rm Z \in \{\rm X, \rm Y\}$) means the state considering pulse correlation and is orthogonal to $\ket{\Psi_{\rm Z}^{\prime\prime}}$. We consider the worst case where $\fid{\Psi_{\rm X}^{\prime\prime\perp}}{\Psi_{\rm Y}^{\prime\prime\perp}}=0$.  Here we only consider the pattern effects limited to the adjacent pulses for simplicity and we define the maximum phase deviation $\overline{\psi}$. Assume we prepare a state $\ket{\alpha}$ and the practical state is $\ket{e^{\rm i\psi}\alpha}$. 
Obviously, $\fid{\alpha}{e^{\rm i\psi}\alpha}=e^{(e^{\rm i\psi}-1)|\alpha|^2}$. According to Eq.~\ref{pattstate}, $\fid{\alpha}{e^{i\psi}\alpha} = \sqrt{1-\epsilon}$. Consequently, we can obtain the relation $\epsilon=1-e^{|\alpha|^2(2{\rm cos}\psi-2)}$. For state $\ket{\alpha}$ after phase modulation ($\theta \in \{0,\pi/2, \pi, 3\pi/2\}$), $\ket{\alpha'} = \ket{e^{\rm i(\psi+\theta)}\alpha}$. Thus the interference results are $\frac{e^{\rm i\psi}\pm1}{\sqrt{2}}\ket{\alpha}$. Consequently, the deviation of intensity in the interference results is $\sin\psi$. This is reflected in the experimental results as the proportion of the deviation of the detection counts for different patterns from their mean values.

We specifically focus on the correlation between adjacent pulses, resulting in a total of sixteen possible patterns. The Client's PM is set to an unmodulated phase 0 and Merchant generates each pattern with an equal probability. 
The pulses interfered at Eve's BS, and the results were measured by detectors $\rm D_1$ and $\rm D_2$.
To reduce the impact of statistical fluctuations and dark counts, we calculated the deviation values of phase 0 and $\pi/2$ by $\rm D_1$'s detection results while the deviation values of phase $\pi$ and $3\pi/2$ are given by $\rm D_2$'s detection results. As presented in Table~S2, the maximum values of $\sin\psi$ are $5.58\times 10^{-3}$, $5.89\times 10^{-3}$, $6.91\times 10^{-3}$ \textcolor{black}{and $7.35\times 10^{-3}$.} Thus, we can get $\psi$, which are equal to corresponding $\sin\psi$ values.

\section*{section S3. Experimental data}
\noindent The calibrated efficiencies of different optical elements in Eve's site are listed in Table~S3. The detection results during implementation are summarized in Table~S4, including the number of all detection events n, the number of detection events under X basis $n_x$,  the number of detection events under Y basis $n_y$ and the number of detection events under different added phases. The number of detection events under different phases is labelled as ``Detected MC", where ``M” (``C”) means an M (C) phase was added on the pulses by Merchant (Client) and detected by $\rm D_i$, i$\in$\{1, 2\}.

\begin{figure*}[ht]
	\includegraphics[width=\textwidth]{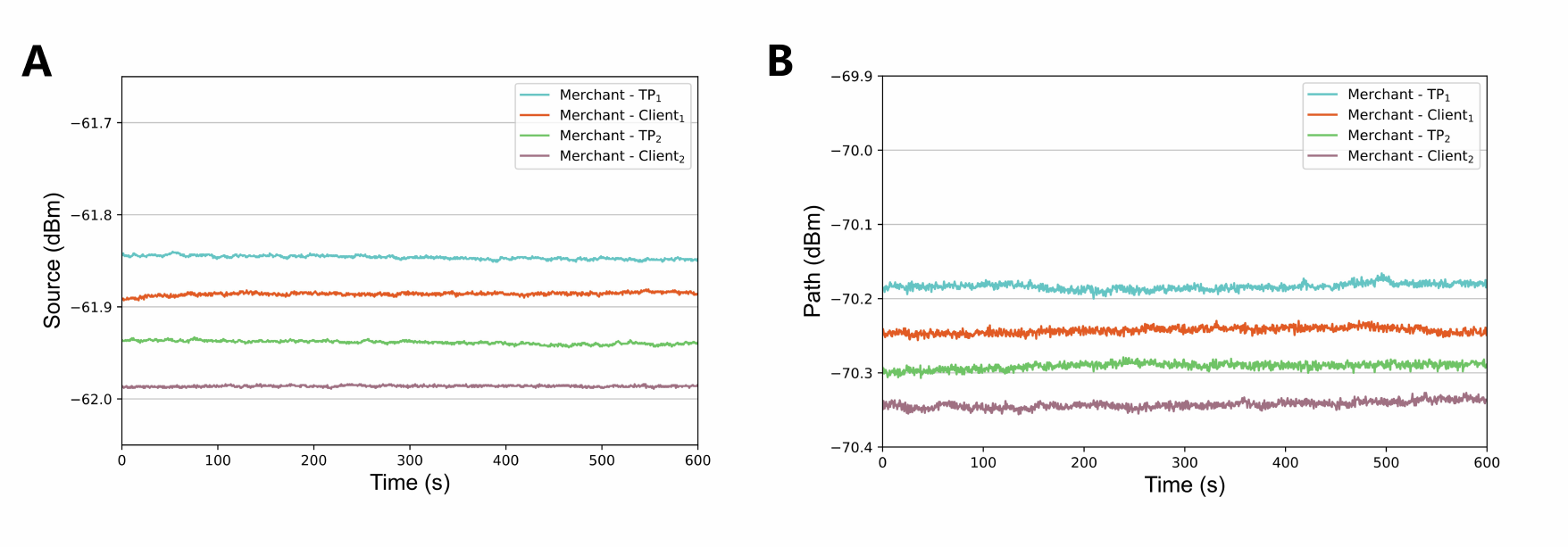}
	\raggedright \textbf{Fig. S1. Optical power fluctuation. (A)} Source fluctuation. \textbf{(B)} Path fluctuation. The maximum fluctuation of the laser source within 100 seconds was measured to be 0.008 dBm, 0.009 dBm, 0.008 dBm, \textcolor{black}{and 0.005 dBm,} respectively. For the path fluctuation, the corresponding values were 0.025 dBm, 0.022 dBm, 0.019 dBm, 0.021 dBm \textcolor{black}{and 0.023 dBm.} As a result, the maximum optical power fluctuations within the 100-second duration were found to be smaller than 0.76\% (0.033 dBm), 0.72\% (0.031 dBm),  0.62\% (0.027 dBm), \textcolor{black}{and 0.65\% (0.028 dBm),} respectively.
    \label{inten}
\end{figure*}

\begin{figure*}[ht]
	\includegraphics[width=\textwidth]{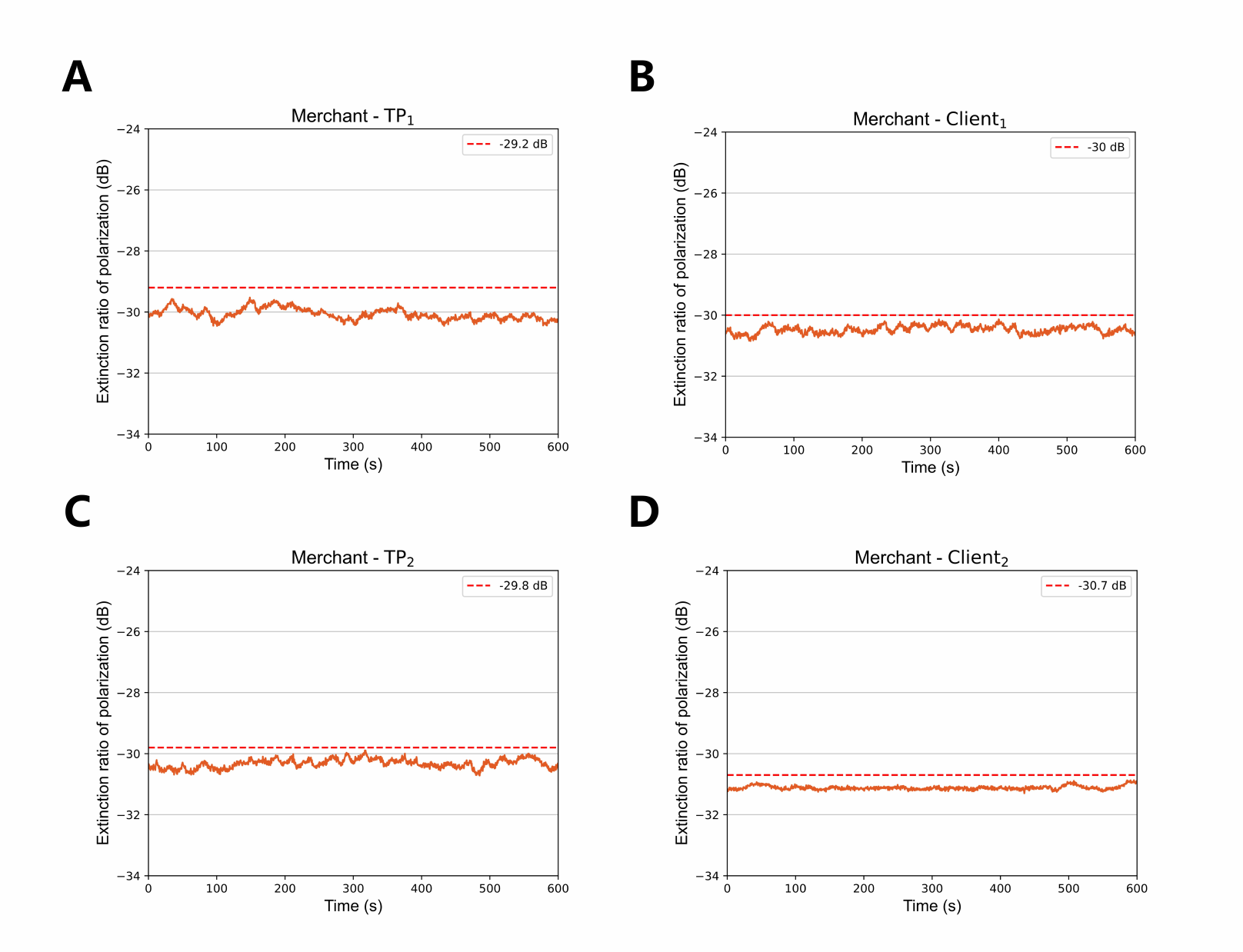}
	\raggedright \textbf{Fig. S2. The extinction ratios of polarization measured over time in four systems.} The obtained values consistently remained below -29.2 dB, -30 dB,  -29.8 dB, \textcolor{black}{and -30.7 dB} during the ten-minute duration.
	\label{pol}
\end{figure*}

\newpage

\begin{table*}[t]
\centering
	
	\textbf{Table S1. Upper bounds of phase shifts under different phases between different pair of participants.} $\rm D_{1,\phi}$ and $\rm D_{2,\phi}$ are the number of clicks detected by $\rm D_1$ and $\rm D_2$ under different phases $\phi$ ($\phi\in \{0, \pi/2, \pi, 3\pi/2\}$) and their upper bounds are denoted by $\overline{\delta}_\phi$. 
\\
\renewcommand\arraystretch{1.2}
	\setlength{\tabcolsep}{0.2cm} 
	\begin{tabular}
		{cccccccccc} \hline \hline 
            &\multicolumn{3}{c}{Merchant - $\rm {TP_1}$}&\multicolumn{3}{c}{Merchant - $\rm Client_1$}&\multicolumn{3}{c}{Merchant - $\rm {TP_2}$} \\ \hline
		$\phi$ & $\rm D_{1,\phi}$ & $\rm D_{2,\phi}$ & $\overline{\delta}_\phi $ & $\rm D_{1,\phi}$ & $\rm D_{2,\phi}$ & $\overline{\delta}_\phi $ & $\rm D_{1,\phi}$ & $\rm D_{2,\phi}$ & $\overline{\delta}_\phi $\\ \hline
		0 & 5399578 &  1386 & -& 5399006 &  1927 & -& 5398064 &  1475 & -\\ 
		$\pi/2$& 335935 & 384598 & $0.013$& 337584 & 383687 & $0.016$& 336621 & 385404 & $0.013$ \\
		
		$\pi$ & 2954 & 6130443 & $0.038$& 3176 & 6135875 & $0.035$& 2817 & 6131470 & $0.035$\\
		
		$3\pi/2$ & 309240 & 340297  & $0.033$& 308194 & 341223 & $0.030$& 308961 & 339748 & $0.033$\\ \hline
		\hline
	\end{tabular}
	\label{tabphase}
\end{table*}

\begin{table*}[ht]
	\renewcommand\arraystretch{1.2}
	
	\setlength{\tabcolsep}{0.2cm} 
	\begin{tabular}
		{cccc} \hline \hline 
            &\multicolumn{3}{c}{\textcolor{black}{Merchant - $\rm {Client_2}$}} \\ \hline
		\textcolor{black}{$\phi$} & \textcolor{black}{$\rm D_{1,\phi}$} & \textcolor{black}{$\rm D_{2,\phi}$} & \textcolor{black}{$\overline{\delta}_\phi$ }\\ \hline
		\textcolor{black}{0} & \textcolor{black}{5390847} &  \textcolor{black}{1175} & \textcolor{black}{-}\\ 
		\textcolor{black}{$\pi/2$}& \textcolor{black}{340889} & \textcolor{black}{380081} & \textcolor{black}{$0.025$} \\		
		\textcolor{black}{$\pi$} & \textcolor{black}{2715} & \textcolor{black}{6163923} & \textcolor{black}{$0.037$}\\		
		\textcolor{black}{$3\pi/2$} & \textcolor{black}{303967} & \textcolor{black}{350989}  & \textcolor{black}{$0.009$}\\ \hline
		\hline
	\end{tabular}
	\label{tabphase}
\end{table*}

\pagebreak

\begin{table*}[ht]
\centering
\renewcommand\arraystretch{1.5}
	 \textbf{Table S2. Detection data for pattern effect.} We recorded the number of clicks of pulses with different phases for each of the sixteen patterns $n_{\rm pa}$, from which the value of $\sin \psi$ can be obtained. We also present $\sin \overline{\psi}$ (the maximum deviation value from the average value of different phases) in the table. \\
	\begin{tabular}        {c@{\hspace{0.64cm}}c@{\hspace{0.64cm}}c@{\hspace{0.64cm}} c@{\hspace{0.64cm}}c@{\hspace{0.64cm}}c@{\hspace{0.64cm}} c@{\hspace{0.64cm}} }  \hline \hline
                    & \multicolumn{2}{c}{Merchant - $\rm {TP_1}$} & \multicolumn{2}{c}{Merchant - $\rm {Client_1}$} &\multicolumn{2}{c}{Merchant - $\rm {TP_2}$} \\ \hline
		Pattern &  $n_{\rm pa}$ &  $\sin \overline{\psi}$ &  $n_{\rm pa}$ &  $\sin \overline{\psi}$ & $n_{\rm pa}$ &  $\sin \overline{\psi}$\\ \hline
		$S_1 \rightarrow S_1$ & 375878 & \multirow{4}{*}{$2.29\times 
            10^{-3}$} 
            & 375953 & \multirow{4}{*}{$1.50\times 10^{-3}$}
            & 375153 & \multirow{4}{*}{$4.69\times 10^{-3}$}\\ 
		$S_2 \rightarrow S_1$ & 374914 &  &376614 & &374603 &\\ 
		$S_3 \rightarrow S_1$ & 375931 &  &375953 & &374532 &\\ 
		$S_4 \rightarrow S_1$ & 376382 &  &375673 & &377112 &\\ \hline
		$S_1 \rightarrow S_2$ & 188037 &\multirow{4}{*}{$3.98\times 10^{-3}$}  & 188555 & \multirow{4}{*}{$3.28\times 10^{-3}$} &187043 & \multirow{4}{*}{$6.91\times 10^{-3}$}\\ 
		$S_2 \rightarrow S_2$ & 187634 &  &188907 & &188196 &\\ 
		$S_3 \rightarrow S_2$ & 189045 &  &188349 & &189378 &\\ 
		$S_4 \rightarrow S_2$ & 188469 &  &187780 & &188763 &\\ \hline
		$S_1 \rightarrow S_3$ & 430459 & \multirow{4}{*}{$3.14\times 10^{-3}$} &429037 & \multirow{4}{*}{$1.52\times 10^{-3}$} &429414 & \multirow{4}{*}{$2.52\times 10^{-3}$}\\ 
		$S_2 \rightarrow S_3$ & 429219 &  &428304 & &428170 &\\ 
		$S_3 \rightarrow S_3$ & 428810 &  &428710 & &427898 &\\ 
		$S_4 \rightarrow S_3$ & 427953 &  &429951 & &427851 &\\ \hline
		$S_1 \rightarrow S_4$ & 214274 & \multirow{4}{*}{$5.58\times 10^{-3}$} &211948 & \multirow{4}{*}{$5.89\times 10^{-3}$} &213956 & \multirow{4}{*}{$6.59\times 10^{-3}$}\\ 
		$S_2 \rightarrow S_4$ & 212405 &  &214372 & &211503 &\\ 
		$S_3 \rightarrow S_4$ & 213151 &  &214151 & &211880 &\\ 
		$S_4 \rightarrow S_4$ & 214556 &  &211993 & &214285 &\\ 
		\hline \hline
	\end{tabular}
	\label{tabpatt}
\end{table*}

\begin{table*}[ht]
\renewcommand\arraystretch{1.5}
	
	\begin{tabular}
        {c@{\hspace{0.64cm}}c@{\hspace{0.64cm}}c@{\hspace{0.64cm}}}  \hline \hline
                    & \multicolumn{2}{c}{\textcolor{black}{Merchant - $\rm {Client_2}$}}\\ \hline
		\textcolor{black}{Pattern} &  \textcolor{black}{$n_{\rm pa}$} &  \textcolor{black}{$\sin \overline{\psi}$} \\ \hline
		\textcolor{black}{$S_1 \rightarrow S_1$} & \textcolor{black}{375521} & \multirow{4}{*}{\textcolor{black}{$2.19\times 10^{-3}$}} \\ 
		\textcolor{black}{$S_2 \rightarrow S_1$} & \textcolor{black}{376223} &  \\ 
		\textcolor{black}{$S_3 \rightarrow S_1$} & \textcolor{black}{375607} &  \\ 
		\textcolor{black}{$S_4 \rightarrow S_1$} & \textcolor{black}{374685} &  \\ \hline
		\textcolor{black}{$S_1 \rightarrow S_2$} & \textcolor{black}{188743} &\multirow{4}{*}{\textcolor{black}{$6.97\times 10^{-3}$}}  \\ 
		\textcolor{black}{$S_2 \rightarrow S_2$} & \textcolor{black}{190131} &\\ 
		\textcolor{black}{$S_3 \rightarrow S_2$} & \textcolor{black}{187795} &\\ 
		\textcolor{black}{$S_4 \rightarrow S_2$} & \textcolor{black}{188588} &\\ \hline
		\textcolor{black}{$S_1 \rightarrow S_3$} & \textcolor{black}{430245} & \multirow{4}{*}{\textcolor{black}{$2.52\times 10^{-3}$}} \\ 
		\textcolor{black}{$S_2 \rightarrow S_3$} & \textcolor{black}{429676} &\\ 
		\textcolor{black}{$S_3 \rightarrow S_3$} & \textcolor{black}{429261} &\\ 
		\textcolor{black}{$S_4 \rightarrow S_3$} & \textcolor{black}{428282} &\\ \hline
		\textcolor{black}{$S_1 \rightarrow S_4$} & \textcolor{black}{215883} & \multirow{4}{*}{\textcolor{black}{$7.35\times 10^{-3}$}}\\ 
		\textcolor{black}{$S_2 \rightarrow S_4$} & \textcolor{black}{213752} &\\ 
		\textcolor{black}{$S_3 \rightarrow S_4$} & \textcolor{black}{215525} &\\ 
		\textcolor{black}{$S_4 \rightarrow S_4$} & \textcolor{black}{216183} &\\ 
		\hline \hline
	\end{tabular}
	\label{tabpatt}
\end{table*}

\begin{table*}[ht]
\centering
\renewcommand\arraystretch{1.5}
	\textbf{Table S3. The efficiencies of optical elements.} The efficiency of Cir 2$\rightarrow$3 means the total efficiency of pulses from circulator's port 2 to port 3. The efficiency of BS-1 means the efficiency of pulses to port 1. The same applies to BS-2.
	\begin{tabular}{c@{\hspace{1.5cm}}c@{\hspace{1.5cm}}}\hline \hline
		Optical element &  Efficiency\\  \hline
		Cir 2$\rightarrow$3  & 86.9\% \\ 
		BS-1  & 85.5\% \\ 
		BS-2  & 86.0\% \\ 
		$\rm{PC_1}$   & 96.2\% \\
		$\rm{PC_2}$   & 95.4\% \\ \hline\hline
	\end{tabular}\label{EFF_OPT}
\end{table*}

\clearpage
\begin{table*}[h!]
\renewcommand\arraystretch{2}
	\textbf{Table S4. Detailed experimental data under different channel losses.}

	\footnotesize\rm
	\begin{tabular*}{\textwidth}{@{}l*{15}{@{\extracolsep{0pt plus 12pt}}l}} 
		\hline\hline
	Channel loss & \multicolumn{2}{c}{15 dB} & \multicolumn{2}{c}{20 dB \hspace{1cm}} & \multicolumn{2}{c}{25 dB \hspace{1cm}} & \multicolumn{2}{c}{\textcolor{black}{20 dB}} \\ \hline
 With fiber spools& \multicolumn{2}{c}{no} & \multicolumn{2}{c}{no}& \multicolumn{2}{c}{no} &\multicolumn{2}{c}{\textcolor{black}{yes}} 
 \\ \hline
	n & \multicolumn{2}{c}{17189504} & \multicolumn{2}{c}{5463449} & \multicolumn{2}{c}{1707401}& \multicolumn{2}{c}{\textcolor{black}{3143582}}\\ 
        $n_x$ & \multicolumn{2}{c}{13919127} & \multicolumn{2}{c}{4424989} & \multicolumn{2}{c}{1382365}& \multicolumn{2}{c}{\textcolor{black}{2545485}} \\ 
        $n_y$ & \multicolumn{2}{c}{189603} & \multicolumn{2}{c}{60019} & \multicolumn{2}{c}{18850}& \multicolumn{2}{c}{\textcolor{black}{35457}} \\ \hline
        Detector & D1 & D2& D1 & D2& D1 & D2& \textcolor{black}{D1} & \textcolor{black}{D2} \\
        Detected 00  &3298168 &1733 &1050551  &523 &328219 &184 &\textcolor{black}{584933} & \textcolor{black}{3099}\\ 
        Detected 0$\pi$ &4570 & 3685584 &1571 & 1171773 &486 &366036&\textcolor{black}{4737} &\textcolor{black}{684026} \\ 
        Detected $\pi$0 &5660 & 3734246 & 1761 &1186466 &463 &371271&\textcolor{black}{4043} & \textcolor{black}{677642}\\ 
        Detected $\pi$ $\pi$ &3186629 & 2537 & 1011575 &769 &315436 & 270 &\textcolor{black}{581509} & \textcolor{black}{5676}\\
        Detected $\frac{\pi}{2}$ $\frac{\pi}{2}$ &34766 & 26 &10931&7 &3425 &1 & \textcolor{black}{9176} &\textcolor{black}{58}\\ 
        Detected $\frac{\pi}{2}$ $\frac{3\pi}{2}$ &22 &59559 &10 & 18708 &6 & 5939 & \textcolor{black}{32} & \textcolor{black}{10537}\\ 
        Detected $\frac{3\pi}{2}$ $\frac{\pi}{2}$ &43 &53998 &11 &17177 & 6   & 5255 & \textcolor{black}{56} &\textcolor{black}{7497}\\ 
        Detected $\frac{3\pi}{2}$ $\frac{3\pi}{2}$ &41141 &48 &13165 &10 & 4212  &6 & \textcolor{black}{8052} &\textcolor{black}{49}\\

		\hline\hline
	\end{tabular*}\label{EXP_DET}
\end{table*}

\end{document}